\documentclass[twoside]{article}
\usepackage{graphicx,times,latexsym,epsfig,amsmath,apeiron}

\begin{document}

\newcommand{\footer}

\setcounter{page}{1}



\title{
 Particle localization and the Notion of Einstein Causality \footnote{{This article appeared in: {\em Extensions of
    Quantum Theory}, 
 Eds. A. Horzela and E. Kapuscik,
 published by Apeiron, Montreal,  
2001, p. 9-16.}}} 
           


\author{
Gerhard C. Hegerfeldt\\
Institut f\"ur Theoretische Physik,
Universit\"at G\"ottingen\\
Bunsenstr. 9,
37073 G\"ottingen, Germany}

\newcommand{\comment}
{The notion of Einstein causality, i.e. the limiting role of the
velocity of light in the transmission of  signals, is discussed.  It is
pointed out that Nimtz and coworkers use the notion of signal velocity
in a different sense
  from Einstein and that their experimental results
are in full agreement with Einstein causality in its ordinary sense.
We also show that under quite general assumptions instantaneous
spreading of particle localization occurs in quantum theory,
relativistic or not, with fields or without. We discuss if this
affects Einstein causality.}
\newcommand{\keywords}{superluminal, signal, localization}
\maketitle
\section{Introduction}
The notion of `Einstein causality'  refers to the
limiting role of the velocity of light in the transmission of signals.
Einstein's principle of finite signal velocity is of
fundamental importance for the foundations of physics, both in
classical as well as in quantum physics. If  signal velocities could
be arbitrarily high this would either lead to the possibility of
absolute clock synchronization and  to a change of special
relativity or to the possible existence of superluminal tachyons with
their associated acausal effects \cite{tachyon}. Hence the name
Einstein causality. 

To be more precise, in this context a {\em signal} means the
experimental creation of any sort of ``disturbance'' at some
space point or small space region and the influence of this on a
measuring device further away. For example, one could produce an
electromagnetic pulse and then measure the field strength at some
other point. The start time of the signal is the time when the
experiment is set into motion, i.e.  when the button is
  pressed. The arrival time of the signal is the {\em first instance} a
measuring device can or does respond. The limiting role of the light velocity
means that the corresponding time difference divided by the distance
cannot exceed $c$.    

 Nimtz and coworkers \cite{Nimtz} have reported superluminal signal
 velocities in tunneling experiments with microwaves.
These  experiments  and their interpretation,
advocated  for example in the article of Nimtz et
al. appearing in this issue, 
 has given rise to considerable controversy 
\cite{Goenner}. It will be shown further below that the controversy is
easily resolved by a careful analysis of the notions used by different
authors. Nimtz and coworkers employ a definition of signal velocity
which is different from the one Einstein had in mind. Using the
old definition it will be seen that the experimental results of Nimtz
and coworkers, sophisticated as they are, do not  contradict  Einstein
causality  in the original sense but,
rather, are in full agreement with it. Thus a conceptual
confusion lies at the heart of the matter which explains a lot of the
controversy.

Are there superluminal phenomena in the quantum realm?
For a free nonrelativistic particle instantaneous spreading of the wave
function is well known.  If, at time $t = 0$, the wave function
vanishes outside some finite region $V$ then the particle is localized
in  $V$ with probability 1. Instantaneous spreading 
implies that the probability of finding the particle arbitrarily
far away from the initial region is nonzero for any $t > 0$. In a
nonrelativistic theory, however, this superluminal propagation  is of
no great concern.

If the localization of a free relativistic particle is
described by the Newton-Wigner position operator then
instantaneous spreading also occurs, as noted in Refs. \cite{Weidlich} and
\cite{Fleming}  (cf. also Ref. \cite{Schlieder}). This also  
 happens for a proposed photon position operator
 \cite{Amrein}. In 1974 the present author \cite{He74} showed that
this phenomenon of instantaneous spreading  is quite general for a
free relativistic particle, 
irrespective of the particular notion of localization, be it in the
sense of Newton-Wigner or others. Later an alternative proof of this
result was given \cite{Skag} and the result was extended to the center-of-mass
motion of relativistic
systems with possibly more than one particle \cite{Perez}.  Ruijsenaars
and the author  
\cite{He80} then showed  that instantaneous spreading occurs for quite
general, relativistic or nonrelativistic, interactions. The main result of
Ref. \cite{He80} was that this instantaneous 
spreading is mainly due to positivity of the energy plus translation 
invariance.  More recently it was shown by the  author \cite{Bohm} that
translation invariance is also not needed.  Hilbert space 
and positivity of the Hamiltonian (energy) suffices to ensure either
instantaneous spreading or, alternatively, confinement in a fixed
region for all times.
Another extension was given by the author \cite{He85} for free
relativistic particles and for 
relativistic systems which have exponentially bounded tails in their
localization outside some region $V$. It was shown that
the state spreads out to infinity faster than allowed by a probability
flow with finite propagation speed. Probably the most astonishing part
of our results is the fact    
that so little is needed to derive them. They hold with and without
field theory and with and without
relativity. Only  Hilbert space and positivity of the energy is
needed. 

What do these results mean for
Einstein causality? This will be discussed in the following where we
concentrate on the role played by positivity of 
the energy for instantaneous spreading. We also briefly discuss
Fermi's two-atom model 
\cite{Fermi,He94}. But first we turn to the Nimtz controversy.

\section{Resolution of the Nimtz controversy}

Nimtz et al. \cite{Nimtz} define in Section 2.2 of their paper in this
issue what
they mean by signal velocity and arrival time. Their definition is
motivated by usage in modern engineering. In particular, their notion
of arrival time is connected to the read-out time of the
signal. However, Einstein had a different meaning in mind 
when he formulated  his principle of the limiting role of the
velocity of light for signal velocities, and this has been explained in the
Introduction.   Definitions 
are of course neither right nor wrong, but clearly the meaning of a
statement  as
well as its truth depend on the  definition of the notions employed in the
formulation of the statement. So what do the Nimtz experiments have to
say on the question of Einstein causality in its original sense? Are
they compatible with it?

In these experiments, typically, a rapid sequence of microwave pulses is
generated. Each pulse is split into two
and sent over  different paths of the same length to a
receiver. Calibration of the path length is achieved by displaying the
two pulse sequences  stroboscopically 
as still pictures on an oscillograph. Then a photonic tunnel barrier
is inserted into one of the paths which
attenuates the corresponding pulses and reshapes them. To compare
tunneled  and non-tunneled pulses  the former  are  re-amplified to
their original amplitude height at the receiving end and again displayed
stroboscopically on the oscillograph. The effect is dramatic. Upon
insertion of the  tunnel barrier the still picture of the
tunneled pulses  makes a jump to earlier times, seemingly
indicating that they are arriving earlier than the non-tunneled pulses.
With the definition of signal velocity and arrival time used by Nimtz
and coworkers this is indeed true.

To see, however, whether this has anything to do with superluminal signal
velocities in 
the Einstein sense it is an eye opener to look at the tunneled pulses
without amplification.  Experimentally
it has been verified by Nimtz and coworkers that the
amplitudes of the tunneled pulses are always {\em below} the amplitude of the
non-tunneled pulses \cite{Aichmann}.  In these experiments,  
 the maxima as well as the half widths of the tunneled pulses are
 ahead of those  of the non-tunneled pulses and therefore arrive earlier. 
This is graphically depicted in Fig. 1 by the pulses traveling from
left to right. The figure is
not to scale and and does not represent 
experimental curves, but is just for illustration.

\begin{figure}
 \epsfig{file=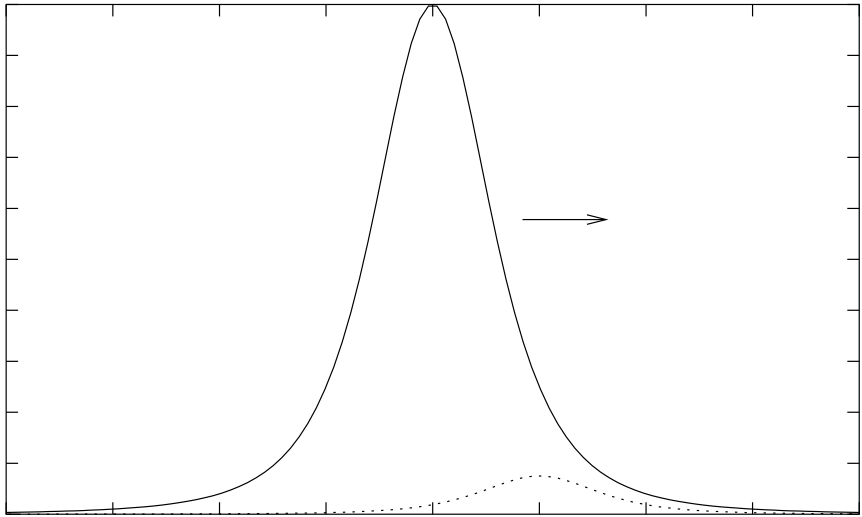}
\caption{Typical behavior  of airborne pulse (solid   line) and tunneled 
  pulse (dashed line), traveling from left to right (not to scale). In the
  experiments, the amplitude of the latter is 
  always smaller than that of the non-tunneled pulse, although its
  maximum arrives at an earlier time.}\label{fig1}
\end{figure}

For the signal velocity in the Einstein sense, however, the arrival time
of the pulse maximum or the read-out time of the half width is not relevant
since they are not used for clock 
synchronization. Relevant, rather, is the first possible response time
of the measuring device, as explained in the Introduction. Now, since
experimentally the tunneled pulses are always below the non-tunneled
pulses in amplitude, any measuring device will 
respond first to the non-tunneled pulses and then to the tunneled
ones, or at most simultaneously to both. Thus the limiting role of the
speed of light as signal velocity in the sense of Einstein is not
violated in the experiments.

What then is superluminal here?  Let us consider the group
and the phase velocity of light. Both are mathematical constructs
useful for the description of electromagnetic phenomena.
It is well known that both can be larger than $c$
\cite{Brillouin}, but this cannot be used for superluminal signals
 in the Einstein sense. Similarly, it has been shown in
Ref. \cite{diplom} that in a somewhat idealized situation the
tunneling pulse can be fully described within 
Maxwell theory by means of another mathematically introduced auxiliary
phase-time velocity notion. Again, this auxiliary
velocity cannot be used for superluminal signal
transmission in the Einstein sense.

So it seems that the controversy about the interpretation of Nimtz's
experiments arises from an indiscriminate use of terminology. Terms
like signal velocity and arrival time are used by Nimtz and
coworkers in a sense different from that  of Einstein.
  Using the notions in the original sense the
 experiments are fully compatible with Einstein  causality as ordinarily
 understood.

\section{Fermi's two-atom model}\label{fermi}
To check the speed of light in quantum
electrodynamics, Fermi \cite{Fermi} considered two atoms, separated by a
distance $R$ and with no photons initially present. One of the atoms
was assumed to be in its ground state, the other in an excited
state. The latter could then decay with the emission of a
photon. Fermi calculated the excitation probability of the atom which
had initially been in its ground state. Using standard approximations he 
found the excitation probability to be zero for $t < R/c$. In
Ref. \cite{He94} the following mathematical result was proved and
applied to the Fermi problem.

{\bf Theorem:} Let $H$ be a self-adjoint operator, positive
or bounded from below, in a Hilbert space ${\cal H}$. For given $\psi_0
\in {\cal H}$ let $\psi_t, t \in I\!\!R$, be defined as 
\begin{equation}\label{t1}
\psi_t = e^{-i Ht} \psi_0~.
\end{equation}
Let $A$ be a positive operator in ${\cal H}, A \ge 0$, and let $p_A(t)$
be defined as
\begin{equation}\label{t2}
p_A(t) = \langle \psi_t , A \psi_t \rangle~.
\end{equation}
Then either
\begin{equation}\label{t3}
p_A(t) \neq 0~~~\mbox{for almost all}~~~ t
\end{equation}
and the set of such $t$'s is dense and open, or
\begin{equation}\label{t4}
p_A(t) \equiv 0~~~\mbox{for all}~~~t~.
\end{equation}

\vspace*{.5cm}
For the proof, which is based on an analyticity argument, both the
positivity of $H$   and of $A$ are needed. Positivity means that all
expectation values of the operator are nonnegative. 
Positivity of $H$ alone is not enough. If A is not positive the
theorem does not hold. In Eq. (\ref{t2}) one can replace $p_A(t)$  by 
$$
p_A(t) = {\rm tr} A  e^{-i Ht}\rho e^{i Ht} 
$$
where $\rho$ is a positive trace-class operator.

If one takes for $\psi_0$ in the theorem the initial state
considered by Fermi and for $A$ the operator describing the
excitation probability, e.g. the projector onto the excited states,
then $p_A(t)$ becomes the excitation probability, and the theorem
states that this probability is immediately nonzero. Already in
\cite{He94} it was discussed how to avoid a possible conflict with
causality, and this was continued in more detail for example in
\cite{He95,BY,LP,MJF}. The conclusion was that the immediate
excitation could be understood in a field-theoretic context through
vacuum fluctuations due to virtual photons. The part of the excitation
due to the second atom behaves causally
\cite{LP,MJF}. Causality then holds for expectation values
after the spontaneous part has been subtracted. This corresponds to
the notion of weak causality, i.e. for expectation values, introduced in
\cite{Schlieder}, which contrasts to the notion of strong causality, i.e. 
causality for
individual events, as discussed in \cite{He95}. Fermi seems to have had
strong causality in mind.
\section{Particle localization and spreading}
Let us suppose that it makes sense to speak  of the probability to
find a particle at a given time inside a space region $V$. This is a
highly nontrivial assumption.
In a quantum theory the probability to find a particle or system
inside $V$ should be given by the expectation of an operator, $N(V)$
say. Since probabilities lie between $0$ and $1$, one must have
\begin{equation}\label{V1}
0 \leq N(V) \le 1~.
\end{equation}

Now let us assume that the system, with state $\psi_0$ at $t = 0$, is
strictly localized in a region $V_0$, i.e. with probability $1$, so
that $\langle \psi_0, N(V_0) \psi_0 \rangle = 1$
or, equivalently,
\begin{equation}\label{V2}
\langle \psi_0, (1 - N(V_0)) \psi_0 \rangle = 0~.
\end{equation}
From Eq. (\ref{V1}) one has $1 - N(V_0) \ge 0$
and hence the theorem can be applied, with 
\begin{equation}
A \equiv 1 -N(V_0). 
\end{equation}
As a consequence one either has
\begin{equation}\label{V4}
\langle \psi_t, N(V_0) \psi_t \rangle \equiv 1~~~{\rm for~all}~~ t
\end{equation}
or
\begin{equation}\label{V5}
\langle \psi_t , N(V_0) \psi_t \rangle < 1~~~{\rm for~almost~all}~~t~.
\end{equation}
The alternative in Eq. (\ref{V4}) means that the particle or system
stays in $V_0$ for all times, as might happen for a bound state in an
external potential. 

Now, if the particle or system is strictly localized in $V_0$ at $t =
0$ it is, {\em a fortiori}, also strictly localized in any larger
region $V$ containing $V_0$. If the boundaries of $V$ and $V_0$ have a
finite distance and  if finite propagation speed would hold then the
probability to find the system in $V$ would also have to be $1$ for
sufficiently small times, e.g.  $0 \le t < \epsilon$. But then the
theorem, with $A \equiv 1 - N(V)$, states that the system stays
in $V$ for {\em all} times. Now we can make $V$ smaller and smaller and let it
approach $V_0$. Thus we conclude that if a particle or system is
strictly localized in a region $V_0$ at time $t = 0$ , then finite
propagation speed implies that it stays in $V_0$ for all times and
therefore prohibits motion to infinity. Or put conversely, if there
exist particle states which are strictly localized in some finite
region at $t = 0$ and later move towards infinity, then finite
propagation speed cannot hold for localization of particles.

This can be formulated somewhat more strongly as follows. If at $t =
0$ a particle is strictly localized in a bounded region $V_0$ then,
unless it remains in $V_0$ for all times, it cannot be
strictly localized in a bounded region $V$, however large, for any 
finite time interval thereafter, and the particle localization immediately
develops infinite "`tails"'. The spreading is over all space except
possibly for "`holes"' which, if any, will persist for all times, by
the same arguments as before. If the theory is translation invariant
then there can be no holes, as shown in Ref. \cite{He80} under some mild
spectrum conditions.

\section{Counterexample Dirac equation?}

At first sight the Dirac equation might seem to be a counterexample to
our results on instantaneous spreading. Indeed, this wave equation is
hyperbolic, implying 
finite propagation speed. For the localization operator $N(V)$ 
one might take  the characteristic function  $\chi_V(\boldsymbol{x})$,
just as in the nonrelativistic case  and in contrast to the
Newton-Wigner operator. Then, for a wave function initially
vanishing  outside a finite region, i.e. of finite support, the
localization does evolve with finite propagation speed! Doesn't this
contradict the results of the preceding section?

This example is instructive since it shows the importance of the  
positive-energy condition. The Dirac equation contains positive and
negative energy states. Now,  consider a solution of the Dirac
equation which vanishes outside some finite region and make the
additional assumption that it
is composed of positive-energy solutions only. Then one gets a contradiction to
our results and therefore the additional assumption must be wrong,
i.e. a solution with
finite support at some time must contain negative-energy
contributions. This means that positive-energy solutions of the Dirac
equation always have {\em 
  infinite} support to begin with! This is phrased as a mathematical
result for instance in the book of Thaller \cite{Thaller}. 

Thus the results of the preceding section do not apply if there are no
strictly localized states in the theory! Strict localization of a
state $\psi$ in a region $V$ 
means that $\langle \psi, N(V) \psi \rangle = 1$, and this gives
\[
0 = \langle \psi, ({\bf 1} - N(V)) \psi \rangle = \| ({\bf 1} -
N(V))^{1/2} \psi \|^2
\]
where the root exists by positivity of $N(V)$. This implies
\begin{equation}\label{14a}
N(V) \psi = \psi .
\end{equation}
Hence $\psi$ is an eigenvector of $N(V)$ for the eigenvalue 1 if
$\psi$ is strictly localized in $V$, and vice versa. The eigenvalue
$0$ means strict localization outside $V$.

The existence or nonexistence of strictly localized states depends on
the form of $N(V)$. For example, if one has a self-adjoint position
operator $\hat{\boldsymbol{X}}$ with commuting components, then $N(V)$ is a
projection operator from the spectral decomposition of
$\hat{\boldsymbol{X}}$
and thus has eigenvalues 1 and $0$. Hence in this case there are
strictly localized states for any region $V$, and the result of the
previous section implies instantaneous spreading.

This instantaneous spreading also occurs for position operators with
self-adjoint but {\em non-}commuting components $\hat{X}_i$. Each
$\hat{X}_i$ has a spectral decomposition whose projection operators
give the localization operators for infinite slabs. Eigenvectors for
the eigenvalue 1 represent states strictly localized in these slabs,
and there is instantaneous spreading in this case, too.

To avoid instantaneous spreading one therefore has to consider
localization operators $N(V)$ which are not projectors, for example
positive operator-valued measures. However, if one insists on
arbitrary good localization, i.e. on tails which drop off arbitrarily
fast, then one runs into our results in Ref. \cite{He85} .

{\em Discussion.} 
Could instantaneous spreading be used for the
transmission of signals if it occurred in the framework of
relativistic one-particle quantum mechanics? 
Let us suppose that at time $t=0$ one could prepare an ensemble of
strictly localized 
(non-interacting) particles by laboratory means, e.g. photons in an
oven. Then one could open a window  and would  observe
some of them at time $t = \varepsilon$ later on the moon. Or to better
 proceed by repetition, suppose one could 
 successively prepare strictly
localized individual particles in the laboratory. Preferably this
should be done with 
different, distinguishable, particles in order to be sure when a
detected particle was originally released. Such a signaling procedure
would have very low efficiency but still could be used for
synchronization of clocks or, for instance, for betting purposes. 

Field-theoretic aspects of our results have been discussed in detail
in Ref. \cite{He98}.  Permanent infinite tails in field theory can  
be understood intuitively through clouds of virtual particles due to
renormalization (`dressed states'). Also,  counters  could be
influenced by vacuum fluctuations.

\end{document}